# New trends in South-South migration: The economic impact of COVID-19 and immigration enforcement[♣]

Roxana Gutiérrez-Romero[1] and Nayeli Salgado[2]

**ABSTRACT**

This paper evaluates the impact of the pandemic and enforcement at the US and Mexican borders on the emigration of Guatemalans during 2017–2020. During this period, the number of crossings from Guatemala fell by 10%, according to the Survey of Migration to the Southern Border of Mexico. Yet, there was a rise of nearly 30% in the number of emigration crossings of male adults travelling with their children. This new trend was partly driven by the recent reduction in the number of children deported from the US. For a one-point reduction in the number of children deported from the US to Guatemalan municipalities, there was an increase of nearly 14 in the number of crossings made by adult males leaving from Guatemala for Mexico; and nearly 0.5 additional crossings made by male adults travelling with their children. However, the surge of emigrants travelling with their children was also driven by the acute economic shock that Guatemala experienced during the pandemic. During this period, air pollution in the analysed Guatemalan municipalities fell by 4%, night light per capita fell by 15%, and homicide rates fell by 40%. Unlike in previous years, emigrants are fleeing poverty rather than violence. Our findings suggest that a reduction in violence alone will not be sufficient to reduce emigration flows from Central America, but that economic recovery is needed.

---

[♣] We acknowledge funding from the Global Challenges Research Fund [RE-CL-2021-01]. We thank the statistical team of the EMIF Sur for answering all our queries about the survey, in particular Víctor Sánchez. We also thank Daniel Acosta Chávez, Tania Rodríguez and David Aban Tamayo for outstanding research assistance. We are grateful to Simetría and its director Alicia Santana Cartas for hosting us in Mexico for a few months to gather data needed for this article.

[1] Corresponding author: Professor of Policy and Quantitative Methods at Queen Mary University of London. Mile End Campus, Bancroft Building, 4th Floor, London, UK, E1 4NS. r.gutierrez@qmul.ac.uk.

[2] Vienna University of Economics and Business, Socioeconomics Department.

# 1. Introduction

The ongoing COVID-19 pandemic has caused unprecedented economic shockwaves and excess mortality, disproportionately affecting several vulnerable groups, such as low-skilled migrant workers (Burton-Jeangros et al., 2020). The international travel bans and lockdowns significantly reduced irregular border crossings in important routes such as the Mediterranean Sea, albeit with heterogenous effects in the region. For instance, Spain experienced an unexpectedly large number of undocumented migrants during the first year of the pandemic.[3] These migrants largely came from developing countries suffering major economic downturns and unemployment rather than conflict (European Commission, 2021). Another similar case is the Mexican-Guatemalan border, a key transit route for Central American migrants. Both Mexico and Guatemala adopted temporary mobility restrictions during the onset of the pandemic, which reduced the number of crossings. Nonetheless, as shown here, there was a rise in the emigration flows of male adults travelling with their children.

The sudden loss of job opportunities and livelihoods for millions of people is an explosive combination for people deciding to migrate seeking opportunities elsewhere despite the risks involved (Makridis & Ohlrogge, 2021). Yet, we still lack a clear understanding of the impacts of the pandemic on local economies and on the most vulnerable, such as people typically displaced by violence or poverty. Has the pandemic increased emigration flows? If so, who is migrating? These are important research questions, particularly in developing countries which experienced simultaneous economic and health shocks amid increased border enforcement. To shed light on these issues, this article examines the economic impact of the pandemic and interior enforcement at the United States (US) and Mexican borders on the emigration flows of Guatemalans during 2017–2020.

This article makes three contributions to the literature. First, we estimate the socio-economic impact of the COVID-19 pandemic on Guatemala using satellite data on night light per capita, pollution levels, and violence as proxied by changes in homicide rates at the municipality level. This analysis helps us to analyse the socio-economic impacts that the country experienced during a period where many national surveys were stopped because of the increased risks of the pandemic. Second, we assess to what extent these socio-economic impacts of the pandemic affected emigration flows from Guatemala. To this end, we use the Survey of Migration to the Southern Border of Mexico, known as EMIF Sur. The EMIF Sur is the largest and most comprehensive survey of migration in Central America. The survey includes information on emigration and immigration flows, as well as those migrants that were sent back by Mexican or US authorities. We focus primarily on the emigration flows from Guatemala. These flows provide representative estimates of the number of crossings, whether documented or undocumented, made by Central American people travelling by land seeking jobs or moving for relocation purposes to Mexico or the US for a month or a more extended period (COLEF, 2013). Roughly 27% of the crossings captured by the EMIF Sur in 2017 were undocumented. This figure increased to 37% in 2020. Thus, the EMIF helps us to shed light on the main drivers of documented and irregular migration. Third, we evaluate whether the number of emigration crossings during 2017–2020 was deterred by the number of Guatemalans deported from Mexico and the number of Guatemalan children deported from the US to each Guatemalan municipality. This is an important research question since North America increased border security precisely to reduce emigration from Central America.

---

[3] For instance, in 2020 the European Union on average experienced a 6-year low in the number of irregular border crossings and a 33% year-on-year decrease in asylum applications. However, Spain, had an increase of 46% in irregular crossings (35,800 migrants) if comparing 2020 to the previous year (European Commission, 2021).

As explained in the next section, the deportation of children and families was the signature of the border enforcement imposed by both the US and Mexico in recent years. Notoriously, in the spring of 2017, the Trump administration directed US border patrol agents to prosecute first-time border crossers and separate undocumented migrant parents from their children (Sieff, 2021). As the family separation policy became public, Trump's administration experienced a fierce backlash, culminating in the US abandoning this policy in June 2018. Soon after, Trump's administration launched negotiations with Mexico to step up the control of irregular migration flows. Border enforcement in the region suffered yet another major shift in 2020. The US reduced the number of apprehensions and deportations of undocumented children and migrant families but increased the number of apprehensions of undocumented migrants travelling on their own. This shift in arrests and deportations, plus the imminent election of Joe Biden, who offered a more humane immigration policy, did not go unnoticed (Aguilar, 2020). Smugglers at the Mexican borders, known as *polleros* or *coyotes*, started encouraging their potential clients to travel with their children to reduce their probability of being arrested or deported (Mcdonnel, 2019).

Given the relevance of migration policy, a substantial literature has analysed the impact of US border enforcement on irregular emigration flows in the region (Amuedo-Dorantes et al., 2015; Espenshade & Acevedo, 1995; Hagan et al., 2008). These earlier studies offer quite mixed and contrasting conclusions. Some studies suggest that apprehensions and deportations from the US effectively deter illegal emigration flows from Central America (Martínez Flores, 2020) but not from Mexico (Hoekstra & Orozco-Aleman, 2021). Other studies suggest that the undocumented migration flows are largely unaffected by the intensity of border enforcement but reduce the intention to remigrate (Espenshade, 1994). In contrast, others find enhanced enforcement and parent-child deportations increase the intention of deportees to remigrate (Amuedo-Dorantes et al., 2015; Amuedo-Dorantes & Pozo, 2014). This earlier literature has primarily analysed the impact of US immigration enforcement *before* our period of analysis. Hence, our study contributes to the literature to understand how emigration flows from Guatemala were impacted by the important border enforcement changes implemented in the US and Mexico.

Our analysis shows that the recent reduction in the number of children deported from the US to Guatemala is associated with the recent rise in emigration flows from Guatemala, particularly of adults travelling with their children. However, the recent surge of emigration crossings was also driven by the economic shocks that local economies suffered in Guatemala during the pandemic. Unlike in previous years, emigrants are not fleeing violence (Clemens, 2021), rather, their families and municipalities experienced substantial income losses and economic inactivity.[4] Our case study can shed light on the experiences of other similar developing countries with legacies of conflict and sudden economic crisis.

**2. Setting and related literature**

Guatemala is an important case study for understanding emigration patterns in the Global South. The country endured a long civil war during 1960–1996 that left it with high levels of poverty (50% living under five dollars a day), an embattled economy, and in the top ten for highest homicide rates worldwide (Roser & Ritchie, 2019). Although market-oriented reforms have gradually provided better macro stability to Guatemala's economy over the last two decades, these gains have not translated into significant economic progress (Meyer,

---

[4] According to a smaller survey carried out in Guatemala and Mexico by the Mixed Migration Centre (2021), over 80% of refugees and migrants had lost income during the pandemic and were increasingly relying on debt to survive.

2021). As Fig. 1 shows, albeit Guatemala's Gross National Income (GNI) per capita has gradually increased (8,870 US dollars in 2019), it remains roughly 45% of the level of Mexico's GNI per capita and just 14% of the GNI per capita in the US. Since Guatemala's southern neighbours, Belize, Honduras and El Salvador, have struggled to address their endemic levels of poverty and have even higher homicide rates, thus a viable route for Guatemalans seeking a way out is heading north, towards Mexico or the US.

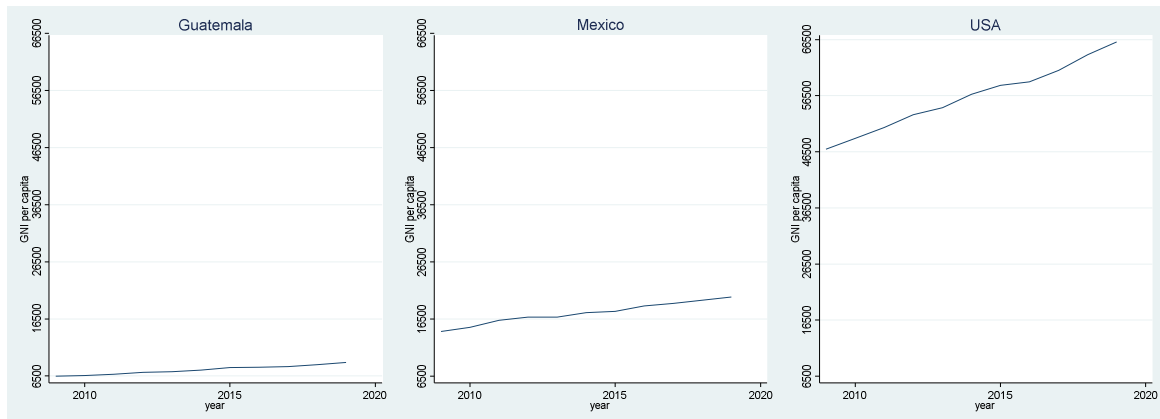

**Fig. 1**. Gross National Income Per Capita, in Current International Dollars and Purchase Power parity, for Guatemala, Mexico and the US, 2004–2019.

The US has traditionally been perceived as the land of opportunities for many Central American migrants. However, as migration policies in the US have hardened and sentiments of anti-migration increased (Kemeny & Cooke, 2017), Mexico has become not only an important crossing path but also a final destination in its own right (COLEF, 2013). According to the EMIF Sur survey analysed in this article, during 2004–2020, nearly 96% of the crossings reported by Guatemalans emigrating north were intended for travelling to Mexico for work purposes and as a final destination. Overall, 31% of respondents during that period claimed they were crossing the border without a valid permit. As Fig. 2 shows, the number of crossings made by Guatemalans emigrating to Mexico or the US rose during the 2000s with clear highs and lows. The EMIF Sur was not conducted during the years 2018 and 2019. However, emigration patterns shifted drastically in 2017. In that year, the number of crossings made by Guatemalans emigrating to Mexico experienced a decline of 65% in 2017, with another decrease of 13% in 2020. The trend in the number of crossings made by adults emigrating with their children to Mexico followed a similar trajectory. However, these flows rose by 27% during the pandemic. Similarly, as shown in Fig. 2, since 2011, the EMIF Sur had recorded zero crossings of adults travelling with their children, claiming they were travelling to the US. However, by 2020, that number increased to 1,452 crossings.

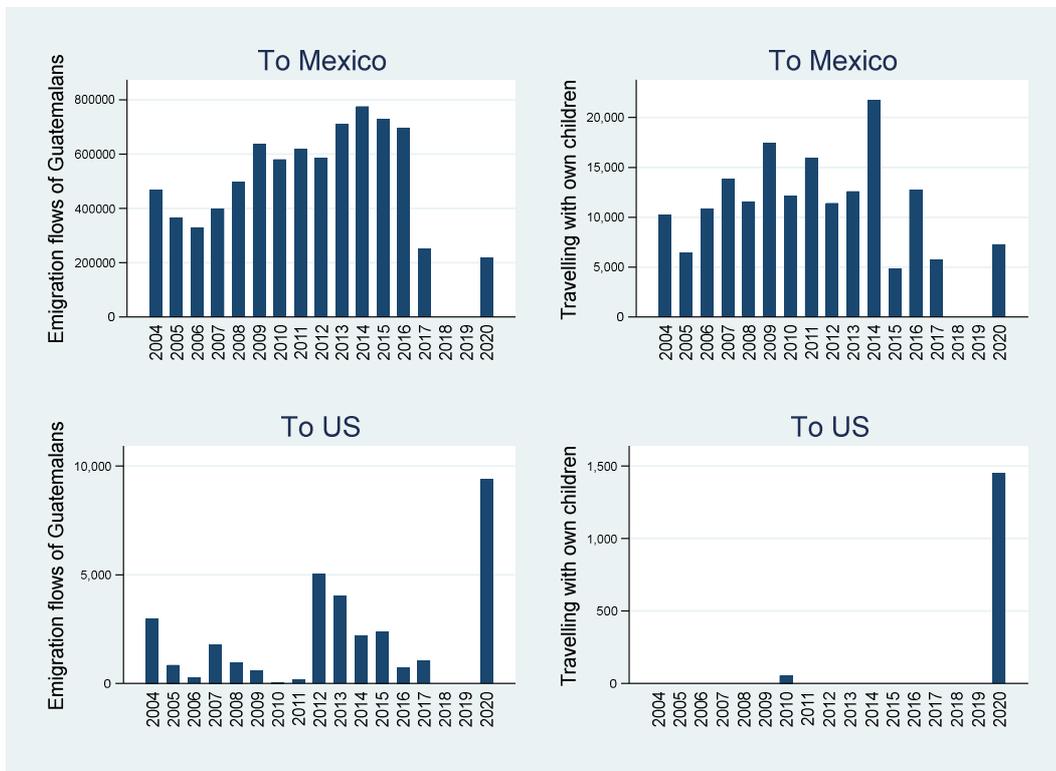

**Fig. 2.** The number of emigration crossings made by Guatemalans leaving for Mexico or the US as the final destination using EMIF Sur and sampling weights. No data are available for 2018–2019.

The EMIF Sur asks migrants the main reason for their most recent crossing. Economic factors were the most critical cited reason before and after the pandemic, with 47% of respondents mentioning lack of employment opportunities and 48% very low income. Less than 3% of migrants cite violence as the main reason for their crossing. Nonetheless, disentangling the net effects of violence and poverty on migration decisions solely from these self-responses is complex. After all, violence and crime can discourage investment, economic activity and limit earning growth (Goulas & Zervoyianni, 2015; REDODEM, 2019). Poverty can also further increase crime and violence (Anser et al., 2020).

In Guatemala, organised crime groups have operated for decades as a result of the counterinsurgents that fought in the civil war and have operated with high levels of impunity. However, violence in Guatemala has been reduced since the country implemented justice reforms in 2007 that improved coordination between law enforcement and prosecutors, significantly reducing impunity and homicide rates (Crisis Group Latin America, 2018). Despite this progress in combating violence, several local gangs, including the *Mara Salvatrucha* and the *M-18*, continue to recruit young adults for lucrative crimes such as extortion and drug smuggling (Swanson & Torres, 2016). Earlier studies, before the pandemic, have found that rises in homicides in Guatemala are associated with more male adults and children emigrating from Guatemala to Mexico (Gutiérrez-Romero, 2022) and more child migrant arrests at the US border (Clemens, 2021).

*2.1 Deportations*

Looking at deportation patterns in Mexico and the US, it is clear that some Guatemalan emigrants will eventually attempt to cross the US border. According to a separate survey module contained in the EMIF Sur, the majority of Guatemalan migrants deported in the Mexican territory, over 90%, claim that they had crossed Mexico for the first time and intended to stay within that country.[5] However, as shown in Fig. 3, the geographical spread of where these deportations occurred in Mexican territory indicates that the US might have been the final intended employment trajectory for some of these Guatemalan migrants. Thus, it is also relevant to analyse the US's border enforcement policy to understand the factors driving emigration flows allegedly destined for Mexico.

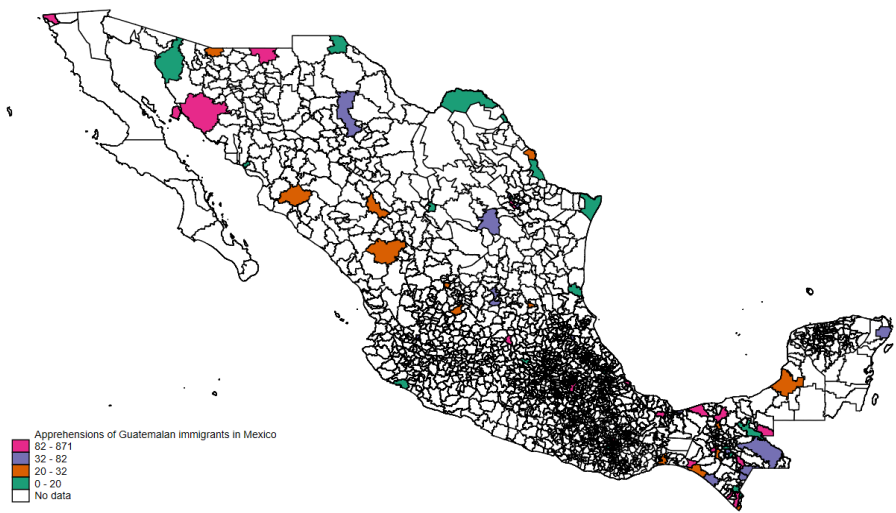

**Fig. 3.** The number of apprehensions of Guatemalan immigrants in Mexico in 2020 using the EMIF Sur survey on returned migrants and sampling weights.

---

[5] The Mexican migration police focuses immigration enforcement in four major migration routes to prevent illegal crossings within territory and to the US (CCINM, 2017).

People's perceptions of migration flexibility and the risk of being denied access if travelling undocumented can be influenced by how easy it is to obtain visas or be deported at the intended destination (Hoekstra & Orozco-Aleman, 2021; Martínez Flores, 2020). The threat of deportation has been an important concern for Guatemalan migrants, particularly during Donald Trump's administration (Abbott, 2022). During his campaign and presidency in the US, Trump claimed that one of his top priorities was to secure the border and prevent illegal immigrants from entering the country, regardless of their age (Shear et al., 2020). Intending to deter irregular crossings at the US-Mexican border, in May 2017, the US government introduced a pilot program in Yuma, Arizona, known as the Criminal Consequence Initiative. This program allowed the prosecution of first-time crossers and the separation of undocumented migrants from their children. About 234 families were separated during July–December 2017, almost the same number of families that were separated in another parallel pilot program (Sieff, 2021). That program, launched in El Paso, Texas, also deliberately and systematically separated undocumented child migrants from their parents at the US border. This policy ensured that apprehended undocumented migrant parents were soon deported whilst their children were left behind under US custody. Due to international and national pressure, Trump ended the family separation policy in June 2018 (Buchanan et al., 2021).

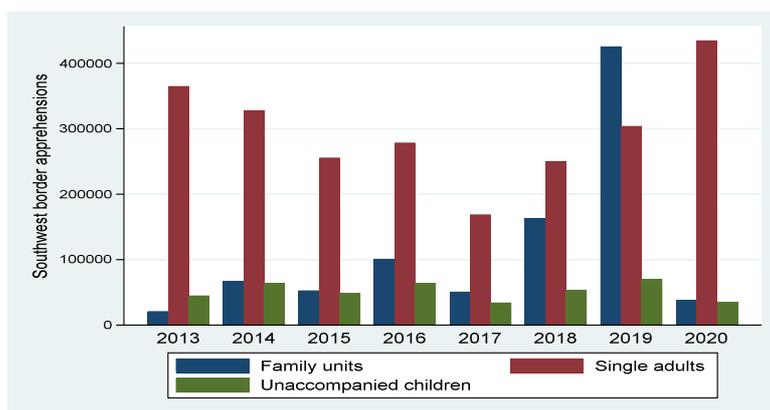

**Fig. 4.** Apprehensions on the US southwest border during 2013–2020 according to US customs and border protection data.

In Fig. 4, we summarise the recent changes in US border enforcement. During Donald Trump's administration, the number of unaccompanied child migrants apprehended at the US-Mexican border gradually increased. From January–December 2017, 34,386 unaccompanied child migrants were arrested. That number continued to grow, reaching a peak in 2019 with 70,418 unaccompanied child migrants apprehended, roughly 40% of them Guatemalan (CBP, 2017). The overall number of unaccompanied child migrants decreased to 35,197 in 2020. Similarly, family unit and single adult migrant apprehensions grew rapidly during 2017–2019. The arrests of family units suddenly fell by 90% in 2020, in contrast to the number of arrests of single adult migrants which continued to rise by another 40% in that year. This contrasting probability of arrest has allegedly prompted smugglers to encourage their clients to travel with their children to reduce deportation risk (Mcdonnel, 2019).

In parallel to the changes in border security in the US, Mexico was forced to increase its border security after Donald Trump's administration threatened to impose heavy trade tariffs on Mexican products and even to walk away from the North American Free Trade Agreement (David, 2018). Under such pressure, in June 2019, Mexico agreed to increase border enforcement. In that month alone, the number of migrants detained and arrested in Mexico increased by 200% (29,000 migrants) and by 180% (22,000 migrants), respectively, the highest monthly migration controls in Mexico in over a decade (Fredrick, 2019). Mexico also agreed to host migrants that had requested asylum in the US until their hearing day at the US immigration court. Nonetheless, between 2019 and 2020, according to official records, the number of Guatemalan migrants deported by Mexican authorities fell by nearly 50% (Secretaría de Gobernación, 2020). During this period, the number of Guatemalan minors deported by Mexican authorities also decreased by 86% (from 12,497 to 1,676) and unaccompanied minors by 13% (from 2,508 to 2,176), according to the Instituto Guatemalteco de Migración (2020). This decrease in deportations in both the US and Mexico was partly driven by the reduction of overall emigration flows, and the sudden mobility restrictions imposed during the first wave of the COVID-19 pandemic.[6]

Earlier studies have shown that Trump's harsh stance on undocumented migration temporarily reduced emigration flows from Central America, particularly at the beginning of his administration (Hoekstra & Orozco-Aleman, 2021). However, since then, the economic impact of the pandemic may have also increased undocumented emigration among certain groups. Undocumented migration flows might have increased if families perceived the reduction of deportations of adults and children from both the US and Mexico as a sign of more lenient immigration policies at northern borders. Next, we empirically assess the impact of the pandemic, and border enforcement on emigration flows from Guatemala.

---

[6] During this period, according to official records of the Ministry of the Interior in Mexico, the number of Guatemalan migrants with work permit in Mexico decreased by 63.3% (going from 9,991 to 3,673). These official records cannot be compared directly to the number of crossings captured by the EMIF Sur.

**3. Data**

*3.1 EMIF Sur*

We use the EMIF Sur, the most comprehensive survey on migration in Central America conducted since 2004. The survey measures various types of migration flows: emigration from Central America, migration from Mexico to Central America, and people returned from Mexican or US authorities to El Salvador, Honduras, and Guatemala. We analyse the survey module that measures emigration flows from Guatemala to Mexico or the US. These flows represent the number of crossings, documented or not, made by people who are emigrating for work, living purposes or visiting family for over a month. The survey follows a probability sampling design for mobile populations (COLEF, 2013).[7] It does not interview people who are migrating for less than a month, are younger than 15 years old, or travelling for other purposes such as tourism or shopping. Migrants are interviewed at the most important border crossings in the Mexican-Guatemalan border, including bus stations and customs inspection points.

We evaluate the impact of immigration policy and socio-economic changes on the emigration flows of Guatemalans at the municipality level. We exclude from the analysis other Central Americans since Guatemalan migrants constitute over 90% of all those interviewed in the EMIF Sur. Also, because in the EMIF Sur it is possible to determine the municipality of residency for Guatemalan migrants only.

Our goal is to assess the impact of migration policies and the COVID-19 pandemic on emigration flows. Thus, we restrict our analysis to the EMIF survey conducted in 2020 and in the previous survey wave, which was conducted in 2017. Table 1 shows why Guatemalans emigrate to Mexico or the US during 2017–2020. We report the number of crossings made by migrants directly interviewed (number of observations unweighted). We also present this information using the sampling weights to get a sense of the overall emigration flows that these crossings represent. We use the sampling weights provided by the EMIF Sur. Most emigrants interviewed are adult males (88%), and most respondents (96%) claimed that they would migrate to Mexico for work reasons. The profile of the emigrants leaving for Mexico is not shown in the table. They are primarily young adult males (average age 32), with no schooling (30%) or just primary school (60%), and a large percentage (40%) speak an indigenous language, Mayan.

---

[7] The survey is administered by *El Colegio de la Frontera*, and various Mexican authorities. For further information about the sampling framework and sampling weights see COLEF (2013) and https://www.colef.mx/emif/bases.html.

**Table 1**

Reasons for emigrating from Guatemala to Mexico or the US in 2017–2020 using EMIF Sur.

| Main reason | All | | | Adult Males | | | Adult Females | | | Children directly interviewed (15-18 years) | | |
|---|---|---|---|---|---|---|---|---|---|---|---|---|
| | Freq unweighted | Freq with sampling weights | Percent with sampling weights | Freq unweighted | Freq with sampling weights | Percent with sampling weights | Freq unweighted | Freq with sampling weights | Percent with sampling weights | Freq unweighted | Freq with sampling weights | Percent with sampling weights |
| To work in Mexico | 7,864 | 445,400 | 95.72 | 6,872 | 372,760 | 96.53 | 819 | 60,605 | 91.73 | 173 | 12,035 | 91.76 |
| To live in Mexico | 22 | 1,621 | 0.35 | 10 | 743 | 0.19 | 12 | 878 | 1.33 | 0 | 0 | 0.00 |
| To visit family or friends in Mexico | 101 | 6,134 | 1.32 | 62 | 4,090 | 1.06 | 35 | 1,799 | 2.72 | 4 | 245 | 1.87 |
| To know Mexico | 27 | 1,737 | 0.37 | 21 | 1,062 | 0.28 | 3 | 351 | 0.53 | 3 | 324 | 2.47 |
| To work in US | 172 | 10,262 | 2.21 | 138 | 7,464 | 1.93 | 31 | 2,348 | 3.55 | 3 | 450 | 3.43 |
| To visit family or friends in US | 4 | 179 | 0.04 | 1 | 27 | 0.01 | 2 | 90 | 0.14 | 1 | 62 | 0.47 |
| Total | 8,190 | 465,333 | | 7,104 | 386,146 | | 902 | 66,071 | | 184 | 13,116 | |

As shown in Table 2, about 2% of adult male migrants going to Mexico are travelling with their children. This figure is higher, 8% for adult females, which nonetheless represents the minority of emigration flows.[8]

**Table 2**

The number of emigration crossings made by migrants leaving Guatemala to Mexico or the US during 2017–2020 using EMIF Sur.

| | Freq unweighted | Freq with sampling weights | Migrated with own children Percent with sampling weights |
|---|---|---|---|
| Men that migrated to Mexico | 6,965 | 378,655 | 2.07 |
| Men that migrated to US | 139 | 7,491 | 7.22 |
| Women that migrated to Mexico | 869 | 63,633 | 8.22 |
| Women that migrated to US | 33 | 2,438 | 37.37 |

*3.2 Deportations from Mexico and the US to Guatemala*

To assess the potential impact of immigration policies in the US and Mexico on emigration flows from Guatemala we use two main indicators. First, we measure the number of people deported from Mexico, regardless of their age or sex, to each Guatemalan municipality.

---

[8] For the years 2017 and 2020, less than 184 children (aged 15–17) were directly interviewed by the EMIF Sur. For the earlier period 2009–2017, there were 3,000 child migrants aged 15–17 directly interviewed by the EMIF Sur.

Second, we also use the number of children deported from the US to each Guatemalan municipality. These statistics are taken from the National Institute of Migration and the National Institute of Statistics in Guatemala.[9]

*3.3 Economic activity, violence, and COVID-19 in Guatemala*

We use two indicators to assess the pandemic's impact on Guatemala's economy: satellite data on night light per capita and pollution, measured by ozone levels. Several other studies have used this type of satellite data to assess changes in economic activity and income levels. This information is particularly relevant for regions with no reliable data at the small-area level or over time during crucial periods such as the pandemic (Bonardi et al., 2021; Sathe et al., 2021). The night light data, measured at the municipality level on an annual basis for 2017 and 2020, was taken from the Earth Observation Group, Payne Institute for Public Policy. The ozone levels, estimated at the municipality level, and at trimester basis for year 2018 and 2020, come from the Sentinel Hub. There are no publicly available ozone data available for the year 2017. Thus for year 2017 we use the levels of ozone for the next availble year, 2018, at trimestre and municipality level.

As Fig. 5 shows, the municipalities with sharper reductions in night light per capita and ozone levels are bordering Mexico and are in the north-centre of Guatemala, which is dedicated to agriculture and has important tourism centres. Fig. 5 also shows the Guatemalan municipalities of the origin of both, adults and children (minors younger than 18) with reported emigration crossings to Mexico in 2017 and 2020. Most emigrants originated from the municipalities near the Mexican border and presumably faced lower migration costs. Nonetheless, in 2020, more emigration flows of adults came from municipalities more scattered in the centre and in the east of Guatemala, closer to El Salvador and Honduras.

To understand the role of violence and the pandemic in driving emigration flows, in Fig. 6, we show the homicide rate, and both the incidence of COVID-19 infection and death rate at municipality level in Guatemala in 2020.[10] Homicide rates are higher in the east of the country. COVID-19 incidence rate is widely spread across the country, but with higher

---

[9] There are no publicly available records on the number of family-child separations for municipality of origin in Guatemala.

[10] The annual homicide rate was obtained from the Guatemalan Police. The COVID-19 rates were taken from the corresponding Ministries of Health.

incidence in the east of the country where more recent emigrants originate.

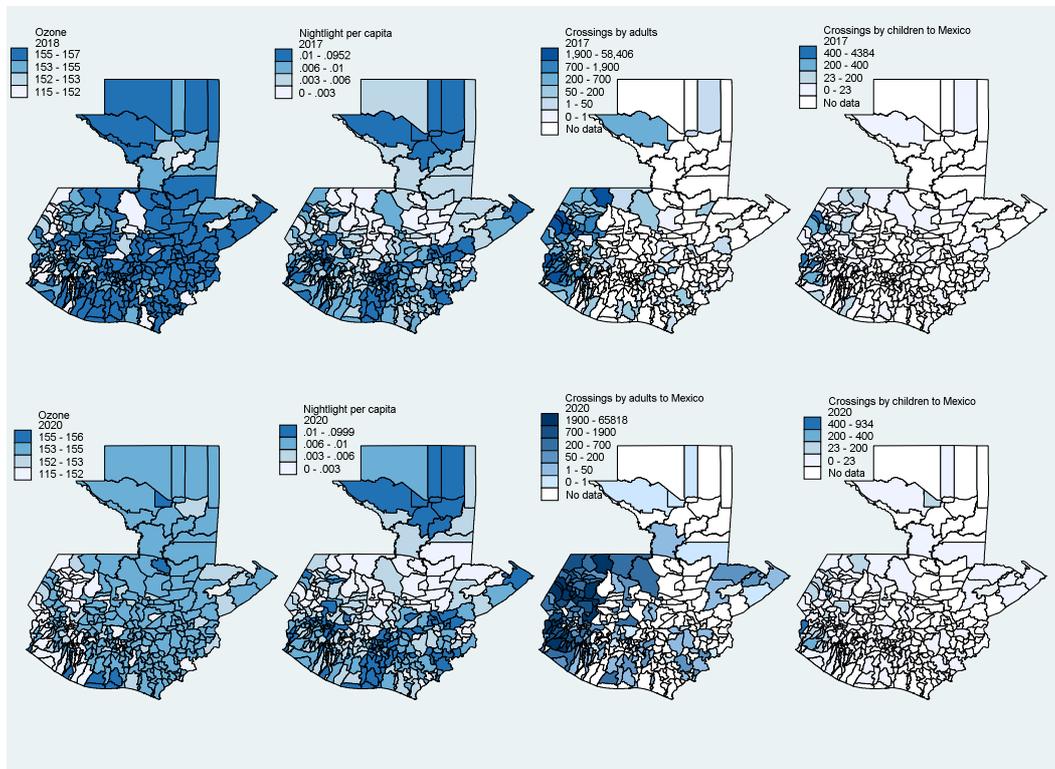

**Fig. 5.** Ozone, night light per capita, and number of emigration crossings from Guatemala to Mexico made by adults and children (calculated with EMIF Sur weights).

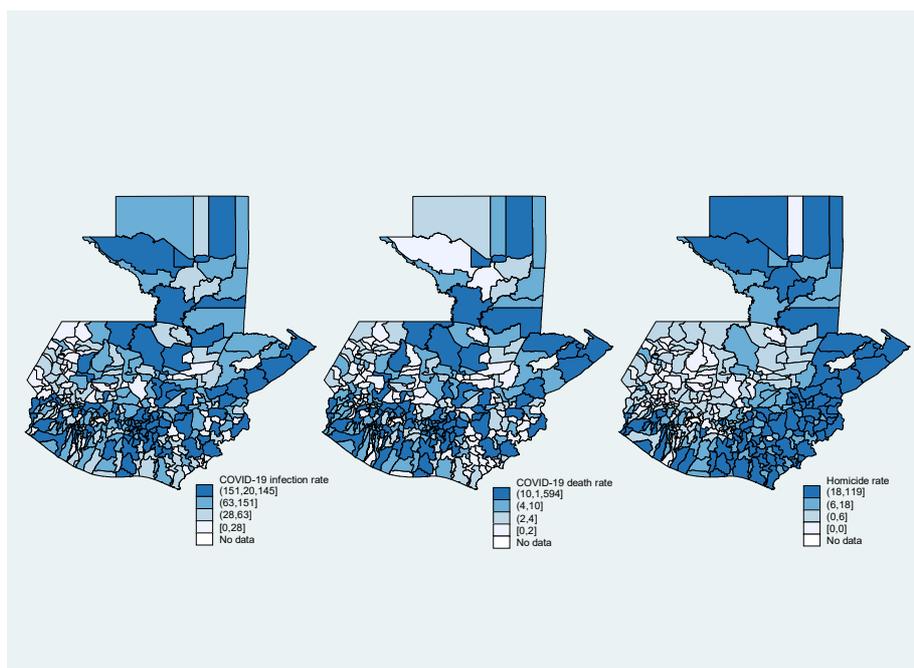

**Fig. 6.** COVID-19 infection, death, and homicide rates in Guatemala 2020.

Most Guatemalans interviewed by the EMIF Sur reported were leaving for Mexico, and specifically for Chiapas, the nearest border state in Mexico. As shown in Fig. 7, COVID-19 infections and associated death rates were lower in Chiapas than in Guatemala. This differential could have motivated some Guatemalans emigrate to Mexico.

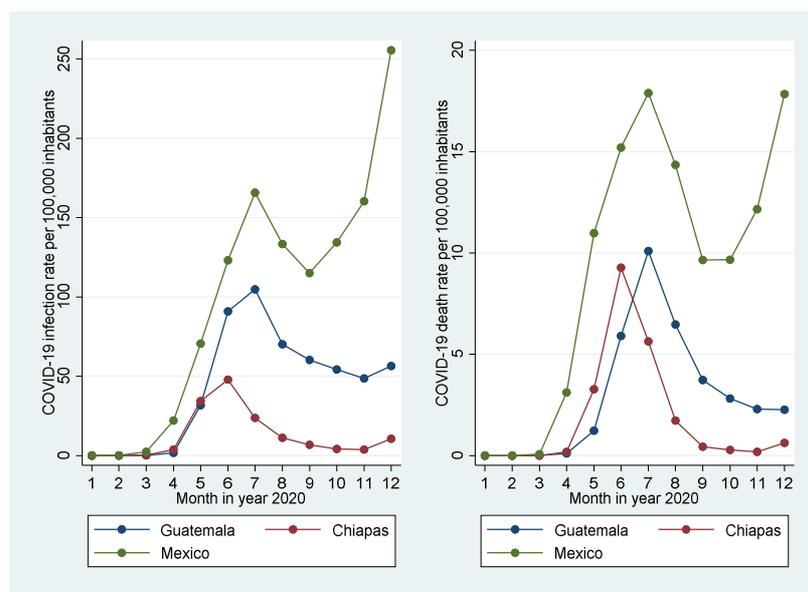

**Fig. 7**. COVID-19 infections and death rate per 100,000 inhabitants in Guatemala, Mexico, and Chiapas.

## 4. Method and results

In this section, we analyse the change in economic activity and violence between 2007 and 2020 in Guatemalan municipalities. Then, we analyse to what extent emigration flows were affected by the recent economic deceleration, along other relevant factors. Since the emigration flows captured by the EMIF Sur are mostly destined for Mexico (98%), in the rest of the paper, we focus only on these flows. We also restrict the analysis to the municipalities for which there is record of emigration flows according to the EMIF Sur. In total there are 340 municipalities in Guatemala. We report the number of municipalities analysed in each of the tables presented.

*4.1 Economic slowdown during the pandemic*

We measure changes in pollution and night light per capita, at municipality level in 2007 and 2020, as shown in equation (1). In this panel fixed effects regression, we also estimate the changes in the homicide rate.

$$Y_{it} = \alpha_0 + \alpha_1\, year_t + \alpha_2\, municipality_i + \varepsilon_{it} \tag{1}$$

where $Y_{it}$ denotes our dependent variable, ozone, night light per capita or homicide rates, all measured for municipality $i$ in time $t$. Ozone varies by trimestre at municipality level, so the data are analysed at those levels. For homicide rates and night light we aggregate at municipality and year levels. As controls we include a dummy variable indicating whether observation belongs to year 2020, and municipality-fixed effects.

**Table 3**

Pollution, night light per capita and homicide rate at the municipality level in Guatemala during 2017–2020.

|  | (1) log ozone | (2) log night light per capita | (3) log homicide rate |
|---|---|---|---|
| Year 2020 (reference year 2017) | -0.040*** | -0.161*** | -0.512*** |
|  | (0.006) | (0.030) | (0.097) |
| Constant | 5.058*** | -4.945*** | 2.634*** |
|  | (0.005) | (0.021) | (0.064) |
| Municipality fixed effects | Yes | Yes | Yes |
| Aggregated at | trimestre level | annual level | annual level |
| Observations | 474 | 282 | 210 |
| Number of municipalities | 184 | 190 | 155 |
| rho | 0.165 | 0.938 | 0.826 |
| R-squared within | 0.0594 | 0.2442 | 0.3417 |
| R-squared between | 0.00923 | 0.0000 | 0.0039 |
| R-squared overall | 0.00182 | 0.0041 | 0.0245 |

*Notes*: rho stands for the interclass correlation, which measures the proportion of variation explained by the individual-specific term. Robust standard errors, clustered at municipality level, in parentheses. Significance levels *** p<0.01, ** p<0.05, * p<0.1

Table 3 shows that the homicide rates in the analysed municipalities decreased by nearly 40%.[11] Also, both ozone and night light per capita fell by approximately 4% and 15%, respectively, between the years 2017 and 2020 in the analysed municipalities experiencing emigration flows. The fall in homicide rates, ozone and night light per capita for the entire country is about 40%, 1.1% and 6.5% respectively for the same period. All these indicators suggest that the pandemic had an important impact on Guatemala's economic activity. Our results support the economic downturn that Guatemala suffered in 2020 according to estimates by the Central Bank of Guatemala. The bank monitors an index of economic activity which reported a decline of 26.6% soon after the lockdown, during the second trimester of 2020, if compared to the previous year. This economic activity index experienced another contraction of economic activity of 4.6% during the third trimester of 2020 and a mild recovery of 3% by the last trimester of that year (Banco de Guatemala, 2021).

---

[11] The dummy regression coefficient of year measures the difference in logs of the dependent variable for the two years being compared. The corresponding change in the dependent variable is obtained by taking the exponential value of the regression coefficient, minus one, and multiplied by 100.

The EMIF Sur survey also provides further evidence of Guatemala's economic downturn in 2020. For instance, in equation (2) we analyse the number of people that worked and provided income in the household of the migrants interviewed in the EMIF Sur, and their employment status before leaving for Mexico. This information is aggregated at the municipality level, which we analyse using panel fixed effects.

$$Labour_{it} = \beta_0 + \beta_1\, year_t + \beta_2\, municipality_i + \beta_3\, X_{it} + u_{it} \qquad (2)$$

where $Labour_{it}$ denotes four dependent variables for municipality $i$ at trimestre $t$ level for year 2017 and 2020. The first one refers to the number of people that work and provide income to the household, including the migrant. The second is the number of migrants that worked in the month before prior departure. Separately we also measure the number of male or female migrants that worked a month before departure. All these dependent variables are aggregated at the municipality level using the sampling weights provided by EMIF Sur. We control for year and municipality-fixed effects and municipality characteristics represented by vector $X$. This vector includes municipalities' COVID-19 infection and death rates, homicide rates, ozone levels and night light. As before, we restrict our analysis to the years 2017 and 2020, for which we have data from the EMIF Sur. The municipalities that have no reported migrants leaving the country in either of the two years are considered with zero reported cases. We cluster the robust standard errors at the municipality level.

Table 4, column 1 and coefficient year, shows that the number of people that worked and provided income to the household, including the migrant, was reduced (by -0.257) between 2017 and 2020. This finding suggests that the migrant's household income was affected, which could have motivated the decision to migrate. In columns (2)–(4), we assess the migrants interviewed by the EMIF Sur (male and female) that worked in the month before departing. The only notable finding is that municipalities with higher COVID-19 death rates

reduced the number of female migrants that had worked previous departing, but not that of male migrants (columns 3 and 4). Specifically, for a unit increase in the COVID-19 death rate in Guatemala, there was a reduction of 0.530 in the number of female migrants that reported working the month before departure.

**Table 4**

Emigrants' employment status before leaving Guatemala during 2017–2020 using EMIF Sur and sampling weights.

|  | (1) Including the migrant, how many people work and provide income to the household | (2) During the last 30 days, migrant worked in place of residency | (3) During the last 30 days, male migrant worked in place of residency | (4) During the last 30 days, female migrant worked in place of residency |
|---|---|---|---|---|
| Covid death rate (at municipality level in Guatemala) | 0.004 | -2.452 | -2.025 | -0.530* |
|  | (0.009) | (1.925) | (1.567) | (0.287) |
| Covid infection rate (at municipality level in Guatemala) | -0.001 | 0.132 | 0.125 | 0.016 |
|  | (0.001) | (0.175) | (0.156) | (0.013) |
| Homicide rate (at municipality level in Guatemala) | 0.007 | -1.531 | -1.536 | -0.194 |
|  | (0.006) | (4.059) | (3.435) | (0.432) |
| log ozone | -1.757* | 30.113 | -87.681 | 86.686 |
|  | (1.006) | (289.014) | (209.213) | (71.803) |
| log night light per capita | 0.296 | 232.188 | 187.776 | 23.546 |
|  | (0.460) | (239.281) | (206.240) | (25.137) |
| Year 2020 (reference year 2017) | -0.257** | 100.234 | 66.960 | 17.299* |
|  | (0.124) | (79.131) | (64.716) | (10.425) |
| Constant | 12.432** | 1,108.960 | 1,474.496* | -317.126 |
|  | (5.609) | (834.636) | (815.932) | (253.138) |
| Municipality fixed effects | Yes | Yes | Yes | Yes |
| Observations | 440 | 451 | 451 | 451 |
| Number of municipalities | 175 | 177 | 177 | 177 |
| rho | 0.485 | 0.529 | 0.466 | 0.276 |
| R-squared within | 0.0594 | 0.0191 | 0.0135 | 0.0279 |
| R-squared between | 0.00923 | 0.0120 | 0.0138 | 0.00688 |
| R-squared overall | 0.00182 | 0.00523 | 0.00662 | 0.000 |

*Notes*: rho stands for the interclass correlation, which measures the proportion of variation explained by the individual-specific term. Robust standard errors, clustered at municipality level, in parentheses. Significance levels *** p<0.01, ** p<0.05, * p<0.1

*4.2 Emigration patterns*

We now analyse the number of emigration crossings made by people leaving Guatemala, all aggregated at the municipality level in Guatemala, and trimestre level in year 2007 and 2020. Emigrants can make several crossings a year. Thus, the number of crossings might not reflect the number of people leaving Guatemala. As such, we refer to these emigration crossings as flows rather than the number of people migrating. These crossings are made by people seeking jobs for a period of a month or longer, whether documented or not. In equation (3), we use panel fixed effects to analyse the impact of changes in border enforcement, COVID-19 incidence, homicide rates and economic slowdown on the number of emigration crossings made from Guatemala for Mexico.

$$Emigration_{it} = \gamma_0 + \gamma_1\, year_t + \gamma_2\, municipality_i + \gamma_3\, D_{it} + v_{it} \tag{3}$$

where $Emigration_{it}$ denotes the number of crossings made by adult emigrants, or those travelling with their children, leaving for Mexico as the final destination for *municipality i* at *year t*. Since emigration flows are primarily driven by males, we separate our dependent variable for male and female migrants. We control for year, and municipality fixed effects as well as by vector *D*. This vector includes Guatemalan municipalities' homicide rates, ozone levels and night light per capita. In addition, vector *D* includes the difference in COVID-19 infections and death rates between municipalities in Guatemala and that of Chiapas. We include this difference as migrants might compare the health risks of emigrating in their municipality and the main point of entry (and allegedly final destination) in Mexico, which for most respondents is Chiapas. As before, we restrict our analysis to the years 2017 and 2020, for which we have data from the EMIF Sur. We also include the number of children that the US deported to each Guatemalan municipality, per trimester, in 2017 and 2020.

**Table 5**

Emigration patterns from Guatemala to Mexico during 2017–2020 using EMIF Sur and sampling weights.

|  | (1) Emigration flows by men leaving for Mexico | (2) Emigration flows by men leaving for Mexico with their | (4) Emigration flows of women leaving for Mexico | (5) Emigration flows of women leaving for Mexico with their | (7) Emigration flows of children leaving for Mexico |
|---|---|---|---|---|---|
| Number of deported children from US to Guatemala | -13.764* | -0.467* | -0.718 | 0.247 | -1.353 |
|  | (8.208) | (0.280) | (2.849) | (0.412) | (1.253) |
| Difference in covid incidence rate between Guatemalan municipality of origin and Chiapas | 0.292 | -0.018 | -0.187 | 0.002 | 0.007 |
|  | (0.455) | (0.014) | (0.156) | (0.014) | (0.022) |
| Difference in covid death rate between Guatemalan municipality of origin and Chiapas | -6.705 | 0.492** | -1.038 | -0.361 | 0.168 |
|  | (4.232) | (0.204) | (2.041) | (0.441) | (0.336) |
| Homicide rate (at municipality level in Guatemala) | -2.609 | -0.543 | -2.463 | -0.580 | 1.804 |
|  | (7.550) | (0.397) | (4.906) | (0.635) | (1.215) |
| log ozone | -1,786.666** | -99.184* | 445.679 | 124.093 | -180.692 |
|  | (799.742) | (53.424) | (564.935) | (94.301) | (139.561) |
| log night light per capita | 777.515 | 41.982 | 302.957 | 58.267 | -34.326 |
|  | (583.372) | (39.328) | (273.606) | (42.798) | (44.753) |
| Year 2020 (reference year 2017) | 171.521 | -0.294 | 120.530 | 11.206 | -11.073 |
|  | (115.385) | (7.513) | (80.556) | (12.073) | (18.104) |
| Constant | 13,647.393** | 740.949** | -605.910 | -318.423 | 750.331 |
|  | (6,617.248) | (319.319) | (1,575.866) | (294.420) | (537.690) |
| Municipality fixed effects | Yes | Yes | Yes | Yes | Yes |
| Observations | 450 | 450 | 450 | 450 | 450 |
| Number of municipalities | 176 | 176 | 176 | 176 | 176 |
| rho | 0.725 | 0.544 | 0.635 | 0.480 | 0.456 |
| R-squared within | 0.0198 | 0.0312 | 0.0335 | 0.0355 | 0.0739 |
| R-squared between | 0.00421 | 0.0389 | 1.56e-05 | 0.00351 | 0.00259 |
| R-squared overall | 0.00106 | 0.0266 | 0.00277 | 9.11e-05 | 0.0143 |

*Notes*: rho stands for the interclass correlation, which measures the proportion of variation explained by the individual-specific term. Robust standard errors, clustered at municipality level, in parentheses. Significance levels *** p<0.01, ** p<0.05, * p<0.1

Table 5, columns (1) and (2), show the regression coefficient of children deported from the US has a negative sign. This means that there is an inverse relationship between these deportations and the number of emigration crossings made by adult males. Specifically,

for a one-point reduction in the number of children deported from the US to Guatemalan municipalities, there is an increase of 13.76 in the number of crossings made by adult males leaving for Mexico and nearly a 0.467 additional crossing made by male adults travelling with their children. However, we find no association between the deportation of Guatemalan children from the US and the emigration flows of women or children leaving for Mexico (Table 5, columns 4-7). This result might again be driven by the fact that the emigration flows analysed are predominantly male-dominated.

We also find that municipalities with a wider difference in COVID-19 death rate than Chiapas experienced higher emigration flows of adult males leaving with their children for Mexico (Table 5, column 2). Moreover, results suggest the municipalities worst affected by the economic slowdown are the ones with the highest emigration flows of male adults leaving for Mexico. For instance, Table 5, columns (1) and (2) show that for a 1% decrease in ozone levels, there was an increase of 1,786 crossings made by male migrants and nearly 100 additional crossings made by male migrants leaving with their children. Again, we find no such effect for female migrants, potentially given the small flows reported of female migrants.

These findings suggest that males migrate to seek better job opportunities, and to potentially reduce their children's health risks. Importantly, we find that homicide rates in Guatemala have no impact on the number of migrants leaving for Mexico, regardless of their age or sex, for the particular period analysed. As mentioned earlier, homicide rates declined by nearly 40% between 2017 and 2020. Thus, perhaps unsurprisingly, violence was not the main driver of the migration flows experienced in 2020.

*4.3 Alternative measures of violence and deportation policy in Mexico*

To assess the robustness of our results, we consider here an alternative measurement of violence. We also test how robust our results are when considering deportations made by Mexican authorities.

In Table A1 in the Appendix, we show two alternative specifications. In the first specification, we use the firearm-homicide rate at the municipality level, which perhaps more directly captures violence triggered by gangs and organised crime common in the region. In the second specification, we add the number of deportees, regardless of age and sex, sent from Mexican authorities to Guatemala. Both specifications provide consistent results with our earlier findings. For instance, Table A1, columns (1) to (4), show that the fewer child migrants are deported from the US, the more crossings are made by adults and adults travelling with their children. Moreover, the number of deportees from Mexico does not affect these emigration flows. This lack of significant association with deportations from Mexico might be related to two aspects. On the one hand, Guatemalan migrants might face a relatively low cost of re-entering Mexico if deported, as opposed to the cost of crossing the US border. On the other hand, roughly 70% of the trips captured by the EMIF Sur from Guatemala to Mexico during our period of analysis are allegedly documented. Also, in line with our earlier findings, violence, captured by firearm-homicide rates at the municipality level in Guatemala, is not related to emigration flows during the period analysed.

*4.4 Addressing endogeneity concerns*

The specifications thus far shown depend on the assumption that US and Mexico border enforcement is largely exogenous and not dependent on the number of emigration crossings that these countries experienced. Other studies have made such an assumption since the change in immigration policy that the US experienced (and subsequently Mexico) was to a large extent driven by the agenda of Donald Trump, who unexpectedly was elected in 2016 (Hoekstra & Orozco-Aleman, 2021). Still, we consider the possibility here that emigration and deportation strategy might be endogenously determined. For instance, one could argue that the number of deportees in the US or Mexico might depend on the number of Guatemalans migrating to Mexico. If so, our key control and the migration flows would be endogenous. Thus, as a robustness check, we use instrumental variables panel fixed effects.

We instrument the potentially endogenous variable, the number of children deported by the US to Guatemala, using two external instruments: the approval of Donald Trump in Texas, one of the US states used by Guatemalans as a main crossing point; and the interaction between this approval rate and the number of homicides committed by firearms, measured at the municipality level in Guatemala for the year 2004. We use Trump's approval rates as an instrument because they are likely to be linked to his administration capacity and willingness to deport child migrants coming from Central America. As mentioned earlier, Trump's administration also was highly influential in Mexico stepping up its border security. Thus this instrument makes it ideal for explaining the number of deportations, whilst not directly affecting the decision of families to migrate. We use the approval rate, at trimestre level for year 2017 and 2019. We use the approval rates for the year 2019, because they may reflect the shift in migration policy in the US, particularly towards family separation and child deportations, and to avoid how the management of the COVID-19 pandemic affected

Trump's approval.[12] We also use the lagged number of homicides committed with firearms in 2004. Over the last two decades, Guatemala has been through a gradual process of disarmament and demobilization that has seen a decline of 18% in the homicide rate between 2005−2017 (Roser and Ritchie, 2019; Naciones Unidas, 2006). Thus, our rationale is that for the municipalities that nearly two decades ago had the highest number of firearm-related homicides could affect migrants' probability of being offered asylum instead of being deported.

The first-stage panel fixed effect IV regressions, in Table A2, show that the instruments are strongly associated with the endogenous regressor, particularly the approval rate of Donald Trump. Table A2 also shows that the F-statistic of the excluded instruments is high (45) and higher than the minimum recommended value of 10. Table A3 shows the second-stage IV panel fixed effects regressions. The bottom rows of Table A3 show that the overidentification and endogeneity tests. According to the Sargan-Hansen overidentification tests, the instruments are uncorrelated with the error term, and correctly excluded from the estimated equation. However, there is no evidence of endogeneity in the IV models ran and no reason why these IV specifications should be preferred to the earleir panel fixed effects.

As another robustness check, we estimate three separate IV panel fixed effects specifications in Table A4. Model 1 uses instead the number of deported migrants from Mexico to Guatemala. This is a more relevant variable since most emigration flows go to Mexico. Deportations from Mexico can also be endogenous; thus we use the same instruments as in our previous specification. In Model 2, we add simultaneously the number of children deported by the US to Guatemala and the number of deportees from Mexico to Guatemala. We instrument these two potentially endogenous variables using the same instruments as before, plus we add the interaction between the number of firearm-related homicides in Guatemala, in year 2004 at municipality level with the approval rate of Donald Trump in Arizona. Arizona is also used by Guatemalans as main crossing point, thus relevant for measuring Trump's approval rates. In Model 3, we drop this third instrument and leave our specification exactly identified with two potential endogenous variables and two instruments.

The first-stage IV panel fixed effects regressions in Table A4 show that three models have relevant instruments, with high F-statistic of excluded instruments (all models above 10). Tables A5 and A6 show the second-stage IV regressions. There are suggestions that the instruments are valid according to the Sargan-Hansen tests. None of the IV models suggest that deportation from the US or Mexico affect emigration flows. However, none of the models have evidence of endogeneity. Thus overall, there are no compelling reasons to prefer the IV specifications over the earlier panel fixed effects specifications.

---

[12] Trump's approval rates were obtained from the Morning Consult Political Intelligence.

## 5. Conclusion

Much of the earlier literature analysing emigration flows from Central America has focused on two key aspects: the role of border enforcement, and violence (Amuedo-Dorantes et al., 2015; Clemens, 2021). In this article, we contributed to that literature by assessing how the COVID-19 pandemic and recent changes in border enforcement affected emigration flows from Guatemala to Mexico during 2017–2020. We used the largest and most comprehensive survey of migrants from Central America, the EMIF Sur. We documented that albeit overall migration crossings from Guatemala were reduced by 10%, there was a rise of nearly 30% in the number of emigration crossings of male adults travelling with their children. Several important changes during the pandemic explain these new trends.

We show that the emigration flows in Guatemala were responsive to the reduction of deportations in the US. For a one-point reduction in the number of children deported from the US to Guatemalan municipalities, there was an increase of nearly 14 in the number of crossings made by adult males leaving from Guatemala for Mexico; and nearly 0.5 additional crossings made by male adults travelling with their children. The intuition behind these findings is that even though most Guatemalan migrants (over 96%) claim they are travelling to work in Mexico as their final destination, deportation patterns within Mexican territory reveal that, eventually, some of these migrants will attempt to cross to the US. We also showed there is no evidence that deportations from Mexico to Guatemala reduce emigration flows, potentially because re-entry to Mexican territory is less costly than attempting to cross to the US.

Another key finding, is that between 2017 and 2020, air pollution in the analysed Guatemalan municipalities fell by 4%, night light per capita fell by aproximately 15%, and homicide rates fell by 40%. These results are consistent with other international studies that have found growth and some crimes were reduced during the pandemic (Bonardi et al., 2021; Nivette et al., 2021). However, the reduction in homicide rates in Guatemala did not contribute to a reduction in emigration flows. We showed that the emigration flows of Guatemalan adults travelling with their children during 2017–2020 were driven by the hardship imposed by the pandemic. For a 1% reduction in pollution, there was an increase of 1,786 crossings made by male emigrants and nearly 100 crossings made by male emigrants leaving with their children from Guatemala to Mexico in search of work and relocation purposes.

These new trends in emigration are worrying because they imply that reducing violence alone will not be enough to stop migration from Central America, but economic recovery and progress are needed.

## Appendix

**Table A1.**

Emigration patterns between Guatemala and Mexico during 2017–2020 using alternative deportation and violence statistics.

|  | (1) | (2) | (3) | (4) | (5) | (6) | (7) | (8) |
|---|---|---|---|---|---|---|---|---|
|  | Emigration flows of men leaving for Mexico | | Emigration flows of men leaving for Mexico with their children | | Emigration flows of women leaving for Mexico | | Emigration flows of women leaving for Mexico with their children | |
| Number of deported children from US to Guatemala | -13.846* | -13.599* | -0.484* | -0.494* | -0.804 | -0.853 | 0.229 | 0.22 |
|  | (8.114) | (8.114) | (0.284) | (0.283) | (2.738) | (2.729) | (0.394) | (0.39) |
| Number of deported migrants from Mexico to Guatemala |  | -0.022 |  | 0.000 |  | -0.022 |  | 0.00 |
|  |  | (0.039) |  | (0.001) |  | (0.013) |  | (0.00) |
| Difference in covid incidence rate between Guatemalan municipality of origin and Chiapas | 0.301 |  | -0.016 |  | -0.175 |  | 0.004 |  |
|  | (0.453) |  | (0.013) |  | (0.147) |  | (0.014) |  |
| Difference in covid death rate between Guatemalan municipality of origin and Chiapas | -6.669 | -1.422 | 0.500** | 0.310** | -1.011 | -0.935 | -0.353 | -0.38 |
|  | (4.186) | (3.887) | (0.211) | (0.150) | (1.937) | (1.923) | (0.433) | (0.44) |
| Firearm-homicides related | -1.093 | -1.322 | -0.291 | -0.279 | -0.373 | -0.235 | -0.308 | -0.31 |
|  | (6.127) | (6.193) | (0.306) | (0.303) | (3.097) | (3.106) | (0.392) | (0.39) |
| log ozone | -1,796.033** | -1,582.987* | -100.782* | -102.738* | 433.220 | 628.481 | 122.366 | 115.3 |
|  | (799.209) | (922.929) | (53.575) | (55.890) | (557.745) | (578.548) | (93.553) | (86.42) |
| log night light per capita | 793.593 | 798.300 | 45.100 | 44.787 | 320.486 | 315.876 | 61.611 | 61.72 |
|  | (604.159) | (603.419) | (40.575) | (40.477) | (300.987) | (299.634) | (46.315) | (46.44) |
| Year 2020 (reference year 2017) | 187.285 | 130.991 | 2.674 | 2.995 | 138.635 | 80.547 | 14.394 | 16.42 |
|  | (120.899) | (159.746) | (7.279) | (7.223) | (102.503) | (100.524) | (14.405) | (18.19) |
| Municipality fixed effects |  |  |  |  |  |  |  |  |
| Constant | 13,744.491** | 12,980.646* | 759.054** | 765.167** | -492.601 | -1,222.445 | -298.962 | -273.0 |
|  | (6,686.991) | (7,143.804) | (322.644) | (328.627) | (1,459.915) | (1,605.595) | (281.293) | (261.9) |
| Observations | 450 | 450 | 450 | 450 | 450 | 450 | 450 | 450 |
| Municipality fixed effects | 176 | 176 | 176 | 176 | 176 | 176 | 176 | 176 |
| rho | 0.727 | 0.729 | 0.566 | 0.565 | 0.646 | 0.654 | 0.506 | 0.50 |
| R-squared within | 0.0306 | 0.0319 | 0.0372 | 0.0367 | 0.0260 | 0.0401 | 0.0337 | 0.034 |
| R-squared between | 0.00942 | 0.0101 | 0.0498 | 0.0495 | 5.95e-07 | 4.38e-05 | 0.00229 | 0.002 |
| R-squared overall | 0.00355 | 0.00375 | 0.0292 | 0.0292 | 0.00199 | 0.00200 | 3.23e-06 | 3.40e- |

*Notes*: rho stands for the interclass correlation, which measures the proportion of variation explained by the individual-specific term. EMIF Sur data weighted with respective sampling weights. Robust standard errors, clustered at municipality level, in parentheses.

Significance levels *** p<0.01, ** p<0.05, * p<0.

**Table A2.**

First-stage IV panel fixed effects corresponding for Table A3: Emigration from Guatemala to Mexico during 2017–2020.

| Endogenous variable: | (1) Number of deported children from USA to Guatemala |
|---|---|
| Donald Trump's approval in Texas x Number of firearm-homicides related by Guatemalan municipalities in 2004 | -0.052*** |
|  | (0.005) |
| Donald Trump's approval in Texas | 0.973*** |
|  | (0.277) |
| Difference in covid incidence rate between Guatemala municipality and Chiapas | 0.002 |
|  | (0.005) |
| Difference in covid death rate between Guatemala municipality and Chiapas | 0.012 |
|  | (0.075) |
| Firearm-homicides related | 0.265** |
|  | (0.119) |
| log ozone | -17.201** |
|  | (8.373) |
| log night light per capita | -6.723 |
|  | (6.770) |
| Year 2020 (reference year 2017) | 16.120*** |
|  | (2.838) |
| Constant | 38.708 |
|  | (52.641) |
| Municipality fixed effects | Yes |
| Observations | 450 |
| Number of municipalities | 0.596 |
| R-squared | 176 |
| F of excluded instruments | 45.14 |
| p-value | 0.00 |
| The Sanderson-Windmeijer (SW) F-of excluded instruments | 45.14 |
| p-value | 0.00 |

*Notes*: Robust standard errors, clustered at municipality level, in parentheses. Significance levels *** p<0.01, ** p<0.05, * p<0.1

**Table A3.**

Second-stage IV panel fixed effects: Emigration patterns between Guatemala and Mexico during 2017–2020.

| | (1) Emigration flows of men leaving for Mexico | (2) Emigration flows of men leaving for Mexico with their children | (3) Emigration flows of women leaving for Mexico | (4) Emigration flows of women leaving for Mexico with their children | (5) Emigration flows of children leaving for Mexico |
|---|---|---|---|---|---|
| Number of deported children from the US to Guatemala | -94.340 | -2.982 | -32.792 | -1.686 | -2.933 |
| | (72.704) | (2.482) | (29.474) | (2.437) | (3.154) |
| Difference in covid incidence rate between Guatemalan municipality of origin and Chiapas | 1.097 | 0.007 | 0.134 | 0.021 | 0.023 |
| | (0.854) | (0.031) | (0.262) | (0.033) | (0.034) |
| Difference in covid death rate between Guatemalan municipality of origin and Chiapas | -10.083 | 0.387 | -2.383 | -0.442 | 0.102 |
| | (8.632) | (0.298) | (3.602) | (0.532) | (0.429) |
| Homicide rate | 3.216 | -0.361 | -0.144 | -0.440 | 1.919 |
| | (19.502) | (0.754) | (8.025) | (0.664) | (1.283) |
| log ozone | -3,059.363** | -138.904** | -60.925 | 93.569 | -205.638 |
| | (1,339.085) | (65.279) | (584.738) | (86.887) | (128.818) |
| log night light per capita | 327.567 | 27.939 | 123.853 | 47.476 | -43.145 |
| | (781.927) | (41.760) | (342.114) | (42.449) | (45.176) |
| Year 2020 (reference year 2017) | 1,281.537 | 34.349 | 562.378 | 37.828 | 10.684 |
| | (1,087.984) | (38.093) | (473.038) | (43.051) | (48.389) |
| Constant | 17,762.318** | 869.374** | 1,032.055 | -219.732 | 830.986* |
| | (7,948.921) | (340.646) | (2,442.729) | (282.061) | (498.367) |
| Municipality fixed effects | Yes | Yes | Yes | Yes | Yes |
| Observations | 450 | 450 | 450 | 450 | 450 |
| Number of municipalities | 176 | 176 | 176 | 176 | 176 |
| Overidentification test | | | | | |
| Sargan-Hansen statistic | 3.11 | 0.00 | 1.95 | 0.40 | 2.07 |
| p-value | 0.08 | 0.95 | 0.16 | 0.53 | 0.15 |
| Weak identification test (Cragg-Donald Wald F statistic): | 45.13 | 45.13 | 45.13 | 45.13 | 45.13 |
| 15% maximal IV size | 11.59 | 11.59 | 11.59 | 11.59 | 11.59 |
| Davidson-Mackinnon test for endogeneity | 0.22 | 0.07 | 0.20 | 1.05 | 3.94 |
| p-value | 0.64 | 0.79 | 0.65 | 0.31 | 0.05 |

*Notes*: Corresponding first-stage IV panel fixed effects regression is in Table A2. EMIF Sur data weighted with respective sampling weights. Robust standard errors, clustered at the municipality level, in parentheses. Significance levels *** p<0.01, ** p<0.05, * p<0.1

**Table A4.**

First-stage IV panel fixed effects for Tables A5 and A6: Emigration from Guatemala to Mexico during 2017–2020 using alternative deportation and violence statistics.

|  | (1) Model 1 | (2) Model 2 | (3) Model 2 | (4) Model 3 | (5) Model 3 |
|---|---|---|---|---|---|
|  | Number of deported migrants from Mexico to Guatemala | Number of deported migrants from Mexico to Guatemala | Number of deported children from USA to Guatemala | Number of deported migrants from Mexico to Guatemala | Number of deported children from USA to Guatemala |
| Donald Trump's approval in Texas x Number of firearm-homicides in Guatemalan municipalities in 2004 | 1.272*** | 1.132** | -0.050*** | 1.272*** | -0.051*** |
|  | (0.364) | (0.479) | (0.005) | (0.364) | (0.005) |
| Donald Trump's approval in Texas | 536.341*** | 522.198*** | 1.031*** | 536.341*** | 0.982*** |
|  | (43.149) | (45.378) | (0.284) | (43.150) | (0.281) |
| Donald Trump's approval in Arizona x Number of firearm-homicides in Guatemalan municipalities in 2004 |  | 1.487* | -0.005 |  |  |
|  |  | (0.843) | (0.008) |  |  |
| Difference in covid incidence rate between Guatemala municipality and Chiapas | 5.827*** | 5.507*** | 0.003 | 5.827*** | 0.002 |
|  | (1.948) | (1.890) | (0.005) | (1.948) | (0.005) |
| Difference in covid death rate between Guatemala municipality and Chiapas | 26.736 | 28.541 | -0.001 | 26.736 | 0.005 |
|  | (23.222) | (22.537) | (0.075) | (23.223) | (0.076) |
| Firearm-homicides related | 3.033 | 1.455 | 0.228** | 3.033 | 0.222* |
|  | (7.048) | (6.919) | (0.115) | (7.048) | (0.116) |
| log ozone | 6,698.442*** | 6,941.878*** | -17.932** | 6,698.442*** | -17.090** |
|  | (1,879.395) | (1,962.629) | (8.377) | (1,879.463) | (8.280) |
| log night light per capita | -244.108 | -196.478 | -8.112 | -244.108 | -7.948 |
|  | (672.453) | (670.693) | (6.737) | (672.477) | (6.742) |
| Year 2020 (reference year 2017) | -1,191.360*** | -1,190.141*** | 15.213*** | -1,191.360*** | 15.218*** |
|  | (230.031) | (230.541) | (2.586) | (230.039) | (2.578) |
| Constant | -50,933.751*** | -52,113.231*** | 36.463 | -50,930.243*** | 32.374 |
|  | (9,014.209) | (9,377.046) | (52.999) | (9,014.646) | (51.840) |
| Year 2020 (reference year 2017) | Yes | Yes | Yes |  |  |
| Observations | 451 | 450 | 450 | 450 | 450 |
| R-squared | 0.500 | 0.502 | 0.590 | 0.500 | 0.589 |
| Number of municipalities | 177 | 176 | 176 | 176 | 176 |
| F-excluded instruments | 14.84 | 10.23 | 28.72 | 85.37 | 42.79 |
| p-value | 0.00 | 0.00 | 0.00 | 0.00 | 0.00 |
| The Sanderson-Windmeijer (SW) F-of excluded instruments | 14.84 | 15.34 | 43.07 | 29.66 | 14.84 |
| p-value | 0.00 | 0.00 | 0.00 | 0.00 | 0.00 |

*Notes*: Corresponding second-stage IV panel fixed effects regressions are in Tables A.5 and A.6. EMIF Sur data weighted with respective sampling weights. Robust standard errors, clustered at municipality level, in parentheses. Significance levels *** p<0.01, ** p<0.05, * p<0.1

**Table A5.**

Second-stage IV panel fixed effects: Emigration patterns between Guatemala and Mexico during 2017–2020 using alternative deportation and violence statistics.

|  | (1) Model 1 | (2) Model 2 | (3) Model 3 | (4) Model 1 | (5) Model 2 | (6) Model 3 | (7) Model 1 | (8) Model 2 | (9) Model 3 |
|---|---|---|---|---|---|---|---|---|---|
|  | Emigration flows of men leaving for Mexico | Emigration flows of men leaving for Mexico | Emigration flows of men leaving for Mexico | Emigration flows of men leaving for Mexico with their children | Emigration flows of men leaving for Mexico with their children | Emigration flows of men leaving for Mexico with their children | Emigration flows of women leaving for Mexico | Emigration flows of women leaving for Mexico | Emigration flows of women leaving for Mexico |
| Number of deported children from US to Guatemala |  | -7.302 | -7.030 |  | -0.211 | -0.202 |  | -1.923 | -1.639 |
|  |  | (5.864) | (5.517) |  | (0.203) | (0.186) |  | (3.000) | (2.737) |
| Number of deported migrants from Mexico to Guatemala | -0.147* | -0.142* | -0.146* | -0.000 | -0.000 | -0.000 | -0.045 | -0.041 | -0.045 |
|  | (0.082) | (0.077) | (0.081) | (0.004) | (0.004) | (0.004) | (0.032) | (0.030) | (0.032) |
| Difference in covid incidence rate between Guatemalan municipality of origin and Chiapas | 1.024* | 1.067* | 1.086* | -0.019 | -0.017 | -0.017 | 0.080 | 0.074 | 0.095 |
|  | (0.561) | (0.597) | (0.601) | (0.030) | (0.030) | (0.030) | (0.123) | (0.119) | (0.129) |
| Difference in covid death rate between Guatemalan municipality of origin and Chiapas | -1.911 | -2.364 | -2.244 | 0.533** | 0.520** | 0.524** | 0.297 | 0.095 | 0.220 |
|  | (6.410) | (6.204) | (6.384) | (0.210) | (0.206) | (0.206) | (2.036) | (2.110) | (2.097) |
| Firearm-homicides related | -0.821 | -0.626 | -0.616 | -0.304 | -0.298 | -0.298 | -0.188 | -0.150 | -0.140 |
|  | (6.261) | (5.952) | (5.959) | (0.311) | (0.311) | (0.311) | (3.070) | (3.013) | (3.008) |
| log ozone | -232.325 | -390.779 | -351.518 | -89.367 | -94.072 | -92.791 | 858.342 | 789.549 | 830.551 |
|  | (988.129) | (956.668) | (985.018) | (60.894) | (59.358) | (60.185) | (723.552) | (664.965) | (696.194) |
| log night light per capita | 853.124 | 809.786 | 810.802 | 47.953 | 46.704 | 46.737 | 317.950 | 307.021 | 308.082 |
|  | (618.381) | (591.156) | (592.048) | (42.314) | (41.866) | (41.887) | (290.754) | (278.264) | (278.679) |
| Year 2020 (reference year 2017) | -394.989* | -284.190 | -298.145 | -4.935 | -1.697 | -2.152 | 6.325 | 43.479 | 28.905 |
|  | (209.926) | (204.010) | (211.635) | (11.313) | (12.198) | (11.836) | (125.092) | (159.245) | (154.141) |
| Constant | 8,055.276 | 8,578.812 | 8,435.332 | 721.069* | 736.665* | 731.984* | -2,067.331 | -1,828.684 | -1,978.526 |
|  | (6,792.603) | (6,752.886) | (6,775.346) | (393.426) | (388.601) | (390.227) | (2,254.932) | (2,080.461) | (2,195.214) |
| Municipality fixed effects | Yes | Yes | Yes | Yes | Yes | Yes | Yes | Yes | Yes |
| Observations | 451 | 450 | 450 | 451 | 450 | 450 | 451 | 450 | 450 |
| Number of municipalities | 177 | 176 | 176 | 177 | 176 | 176 | 177 | 176 | 176 |
| Overidentification and endogeneity tests |  |  |  |  |  |  |  |  |  |
| Sargan-Hansen statistic | 1.32 | 0.42 | 0.00 exactly | 1.29 | 0.36 | 0.00 exactly | 0.43 | 1.55 | 0.00 exactly |
| p-value | 0.25 | 0.52 | identified | 0.26 | 0.55 | identified | 0.51 | 0.21 | identified |
| Weak identification test (Cragg-Donald Wald F statistic): | 14.84 | 10.23 | 14.83 | 14.84 | 10.23 | 14.83 | 14.84 | 10.23 | 14.83 |
| 10% maximal IV size | 11.59 | 8.18 | 4.58 | 11.59 | 8.18 | 4.58 | 11.59 | 8.18 | 4.58 |
| Davidson-Mackinnon test for endogeneity | 1.55 | 0.88 | 0.89 | 0.02 | 0.11 | 0.12 | 0.56 | 0.23 | 0.30 |
| p-value | 0.21 | 0.42 | 0.41 | 0.88 | 0.90 | 0.89 | 0.45 | 0.79 | 0.74 |

*Notes*: Corresponding first-stage IV panel fixed effects regressions are in Table A4. EMIF Sur data weighted with respective sampling weights. Robust standard errors, clustered at municipality level, in parentheses. Significance levels *** p<0.01, ** p<0.05, * p<0.1

**Table A6.**

Second-stage IV panel fixed effects: Emigration patterns of children, and women leaving with their children between Guatemala and Mexico during 2017–2020 using alternative deportation and violence statistics.

| | (1) Model 1 Emigration flows of women leaving for Mexico with their children | (2) Model 2 Emigration flows of women leaving for Mexico with their children | (3) Model 3 Emigration flows of women leaving for Mexico with their children | (4) Model 1 Emigration flows of children leaving for Mexico | (5) Model 2 Emigration flows of children leaving for Mexico | (6) Model 3 Emigration flows of children leaving for Mexico |
|---|---|---|---|---|---|---|
| Number of deported children from US to Guatemala | | -0.346 | -0.315 | | 0.233 | 0.269 |
| | | (0.441) | (0.410) | | (0.493) | (0.445) |
| Number of deported migrants from Mexico to Guatemala | 0.003 | 0.003 | 0.003 | -0.006 | -0.005 | -0.006 |
| | (0.004) | (0.004) | (0.004) | (0.004) | (0.004) | (0.004) |
| Difference in covid incidence rate between Guatemalan municipality of origin and Chiapas | -0.009 | -0.008 | -0.006 | 0.025 | 0.020 | 0.023 |
| | (0.031) | (0.033) | (0.033) | (0.043) | (0.044) | (0.042) |
| Difference in covid death rate between Guatemalan municipality of origin and Chiapas | -0.434 | -0.462 | -0.449 | 0.354 | 0.351 | 0.367 |
| | (0.453) | (0.473) | (0.465) | (0.375) | (0.373) | (0.380) |
| Firearm-homicides related | -0.313 | -0.304 | -0.303 | 1.412 | 1.403 | 1.404 |
| | (0.386) | (0.385) | (0.383) | (0.975) | (0.984) | (0.987) |
| log ozone | 95.688 | 85.867 | 90.341 | -105.372 | -105.986 | -100.812 |
| | (87.064) | (80.563) | (83.647) | (128.344) | (122.713) | (125.743) |
| log night light per capita | 60.639 | 58.624 | 58.740 | -36.091 | -34.605 | -34.471 |
| | (45.253) | (44.443) | (44.444) | (48.012) | (46.928) | (47.077) |
| Year 2020 (reference year 2017) | 24.268 | 30.202 | 28.612 | -51.535* | -53.402* | -55.240* |
| | (17.735) | (21.672) | (20.800) | (27.066) | (31.230) | (30.389) |
| Constant | -201.962 | -168.623 | -184.976 | 447.776 | 452.251 | 433.344 |
| | (272.529) | (253.459) | (263.972) | (497.819) | (481.791) | (494.790) |
| Municipality fixed effects | Yes | Yes | Yes | Yes | Yes | Yes |
| Observations | 451 | 450 | 450 | 451 | 450 | 450 |
| Number of municipalities | 177 | 176 | 176 | 177 | 176 | 176 |
| Overidentification and endogeneity tests | | | | | | |
| Sargan-Hansen statistic | 0.98 | 1.89 | 0.00 | 0.20 | 0.42 | 0.00 |
| p-value | 0.32 | 0.17 | exactly identified | 0.66 | 0.52 | exactly identified |
| Weak identification test (Cragg-Donald Wald F statistic): | 14.84 | 10.23 | 14.83 | 14.84 | 10.23 | 14.83 |
| 10% maximal IV size | 11.59 | 8.18 | 4.58 | 11.59 | 8.18 | 4.58 |
| Davidson-Mackinnon test for endogeneity | 0.12 | 0.48 | 0.39 | 0.15 | 1.47 | 1.53 |
| p-value | 0.73 | 0.62 | 0.68 | 0.70 | 0.23 | 0.22 |

*Notes*: Corresponding first-stage IV panel fixed effects regressions are in Table A4. EMIF Sur data weighted with respective sampling weights. Robust standard errors, clustered at municipality level, in parentheses. Significance levels *** p<0.01, ** p<0.05, * p<0.1